\documentclass[twocolumn,aps,preprint,letter,10pt]{revtex4}
\usepackage{graphicx,graphics,float}% Include figure files
\usepackage{dcolumn}% Align table columns on decimal point
\usepackage{bm}% bold math
\usepackage{amsmath,amssymb}
\usepackage{mathrsfs}
\begin{document}
\title{Condensation driven by a quantum phase transition}
\author{Miguel Alvarez}\email{migue_leo93@hotmail.com}
\author{Jose Reslen}\email{reslenjo@yahoo.com}
\affiliation{Coordinaci\'on de F\'{\i}sica, Universidad del Atl\'antico,
Carrera 30 N\'umero 8-49, Puerto Colombia.}%
%
%\date{}
%\date{\today}

\newcommand{\Keywords}[1]{\par\noindent{\small{\em Keywords\/}: #1}}

\begin{abstract} 
The grand canonical thermodynamics of a bosonic system is studied in order to
identify the footprint of its own high-density quantum phase transition. The phases
displayed by the system at zero temperature establish recognizable patterns at
finite temperature that emerged in the proximity of the boundary of the equilibrium
diagram. The gaped phase induces a state of collectivism/condensation at finite
temperature in which population cumulates into the ground state in spite of
interacting attractively. The work sets the foundation to approach the effect of
attraction in the formation of a molecular condensate.\\
\end{abstract}
\maketitle
\section{Introduction}
The understanding of collectivism is a crucial task in the search for practical
applications where a number of phases taking place in the low-energy range could be
utilized in mass-produced technology.  The process by which a large number of
particles, in particular bosons, collectively occupy the same single-body state
gives rise to coherent phases where the features of a single state magnify and
therefore manifest at the macroscopic level. The archetype of collectivism is
Bose-Einstein condensation \cite{bec}, a mechanism that constitutes the
interpretative ground of other phenomena like superfluidity or superconductivity.
In essence, condensation is driven by statistical effects at the single-body
level, but interaction cannot be completely suppressed in any practical scenario.
The role of attractive interaction, in particular, has proved detrimental in a
number of studies
\cite{gerton,smith,shi,kim,ueda,stoof,huang,bishop,dodd,berge,carr}, but
experimental evidence of condensation in attractive systems has been reported
\cite{hulet}. In this context there must be a range  over which collectivism can be
sustained under the effect of interaction, especially when the interaction role can
be captured using single-body terms. The relevance of this phenomenon is
notorious in times when the first prototypes of quantum computation have
demonstrated quantum supremacy \cite{qsnature,qsscience} and control mechanisms are
increasingly necessary to extent the current capabilities. This has boosted the
interest in the realization of novel collective phases such as the molecular
condensate \cite{chin,warner}, which in turn evidences the need of a better
understanding of the process of condensation under attractive fields since it
may prove a precursor of molecular formation. 

Let us consider a system of one-species bosons that can tunnel between two
equal-energy wells \cite{sipe}. Bosons can interact among them only when they
occupy the same well.  This interaction is attractive, so it tends to pack
bosons together. The system is modeled via the next quantized Hamiltonian
{\footnotesize
\begin{gather}
\hat{H}_M = \delta (\hat{a}_1^{\dagger}\hat{a}_2 + \hat{a}_2^{\dagger} \hat{a}_1) - i \gamma
(\hat{a}_1^{\dagger}\hat{a}_2 - \hat{a}_2^{\dagger} \hat{a}_1)- \frac{\lambda}{M}
(\hat{m}_1^2 + \hat{m}_2^2).
\label{e04091}
\end{gather}
}
Ladder operators obey standard bosonic commutation rules,
$[\hat{a}_1,\hat{a}_1^\dagger] = [\hat{a}_2,\hat{a}_2^\dagger] = 1, \text{ }
[\hat{a}_1,\hat{a}_2] = 0$. The number operators 
\begin{gather}
\hat{m}_1 = \hat{a}_1^\dagger \hat{a}_1 \text{, }  \hat{m}_2 = \hat{a}_2^\dagger \hat{a}_2,
\end{gather}
determine the particle occupation at each well. Integer $M$ is the total number of
particles in the system, so that $M = \hat{m}_1 + \hat{m}_2$. Constants $\gamma$
and $\lambda$ modulate the intensity of hopping and interaction respectively. Both
are considered strictly positive in this work. Constant $\delta$ modulates the
intensity of a hopping process in which the phase of one well with respect to the
other is unchanged.  Henceforth the  energy-scale is chosen so that
$\delta=1$.  By definition $\hat{H}_0=0$.  Hamiltonian (\ref{e04091}) has the
essential elements of a Bose-Hubbard model, which has been studied from the
perspective of the grand-canonical formalism as a descriptor of the transition
between Mott insulator and superfluid \cite{fisher} and also verified
experimentally with ultracold atoms \cite{greiner,atala}. The main difference with
the current study is that here the interaction is attractive and has been scaled
with respect to the number of particles. The latter feature can also be seen in
related models such as the Lipkin-Meshkov-Glick model and the infinite-range Ising
model, from which Hamiltonian (\ref{e04091}) can be obtained as a second
quantization
\cite{jimenez}.

Assuming that to first order the ground state adopts a {\it collective}
form, it can be written as 
\begin{gather}
|G (\theta,\varphi) \rangle = \frac{\left. {\hat{b}_1^{\dagger}}\right.^M
|0,0\rangle}{\sqrt{M!}},\text{ } \hat{b}_1^{\dagger} =  \hat{a}_1^{\dagger} \cos
\theta-  \hat{a}_2^{\dagger} e^{i \varphi}\sin \theta,
\end{gather}
so that $[\hat{b}_1,\hat{b}_1^\dagger]=1$. The angle domains are
$0\le\theta\le \pi/2$ and $0\le\varphi<2\pi$, ensuring that different
angle pairs correspond to genuinely different modes. In order to find the correct
angles, the energy function
\begin{gather}
E(\theta,\varphi) = \langle G (\theta,\varphi) | \hat{H}_M  G (\theta,\varphi)
\rangle,
\end{gather}
is minimized over $\theta$ and $\varphi$. From the procedure shown in appendix
\ref{a04191} it follows that the system undergoes two distinct phases determined by
the value of $\lambda$ relative to the critical value $\lambda_c =
\sqrt{1+\gamma^2}$. In the region $\lambda \le \lambda_c$ the energy displays a
single minimum at
\begin{gather}
\theta^\star = \frac{\pi}{4}, \text{ } \cos \varphi^\star = \frac{1}{\lambda_c},
\text{ } \sin \varphi^\star = \frac{\gamma}{\lambda_c}.
\label{e04071}
\end{gather}
The corresponding ground state energy is
\begin{gather}
E_{\lambda \le \lambda_c}^{\star} = -M \left( \lambda_c + \frac{\lambda}{2} \right).
\end{gather}
From quantum theory it is known that because in this phase the
ground state is non-degenerate it must display the Hamiltonian's symmetries.
Hamiltonian (\ref{e04091}) commutes with the antiunitary operator composed by the
swapping of subscripts $1 \leftrightarrow 2$ followed by complex conjugation. As a
consequence, the number of particles on each well must be the same. 

In the region $\lambda \ge \lambda_c$ the energy displays two equal minima located
at $(\theta_1^\star,\varphi^\star)$ and $(\theta_2^\star,\varphi^\star)$. Here
$\varphi^\star$ is the same than in equation (\ref{e04071}). In addition
\begin{gather}
\theta_1^\star = \frac{1}{2} \arcsin \frac{\lambda_c}{\lambda}, \text{ } \theta_2^\star =  \frac{\pi}{2} - \theta_1^\star.
\end{gather}
Notice that by making $\theta$ different from $\pi/4$ the balance of
occupation between the wells is broken and the proposed solutions do not
independently display the Hamiltonian's inversion symmetry. The corresponding
ground state energy is
\begin{gather} 
E_{\lambda \ge \lambda_c}^\star = -M \left( \lambda + \frac{\lambda_c^2}{2 \lambda}
\right ).  
\end{gather}
The phase change is rooted in a change in the Hamiltonian spectrum, which goes from
non-degenerate for $\lambda < \lambda_c$ to two-fold degenerate for $\lambda \ge
\lambda_c$. In the latter case the equilibrium configuration corresponds to a
maximally mixed state in the space spanned by two linearly independent ground
states. However, it is often argued that small perturbations tip the scale to
either of the constituent pure-states, such that the phase change is equivalent to
a continuous break of symmetry of a pure ground-state as a function of the
Hamiltonian's parameters. This process is known as a quantum phase transition and
takes place at zero temperature since it is in this instance that the equilibrium
state coincides with the ground state. Here it is intended to show that
under specific conditions this quantum phase transition can manifest at finite
temperature, although not as a symmetry breaking of a quantum state in
pure form but as a change in the collective behavior of a grand canonical ensemble
of particles.
\section{Thermodynamic state of the open system}
\begin{figure*}
\begin{center}
\includegraphics[width=0.35\textwidth,angle=-90]{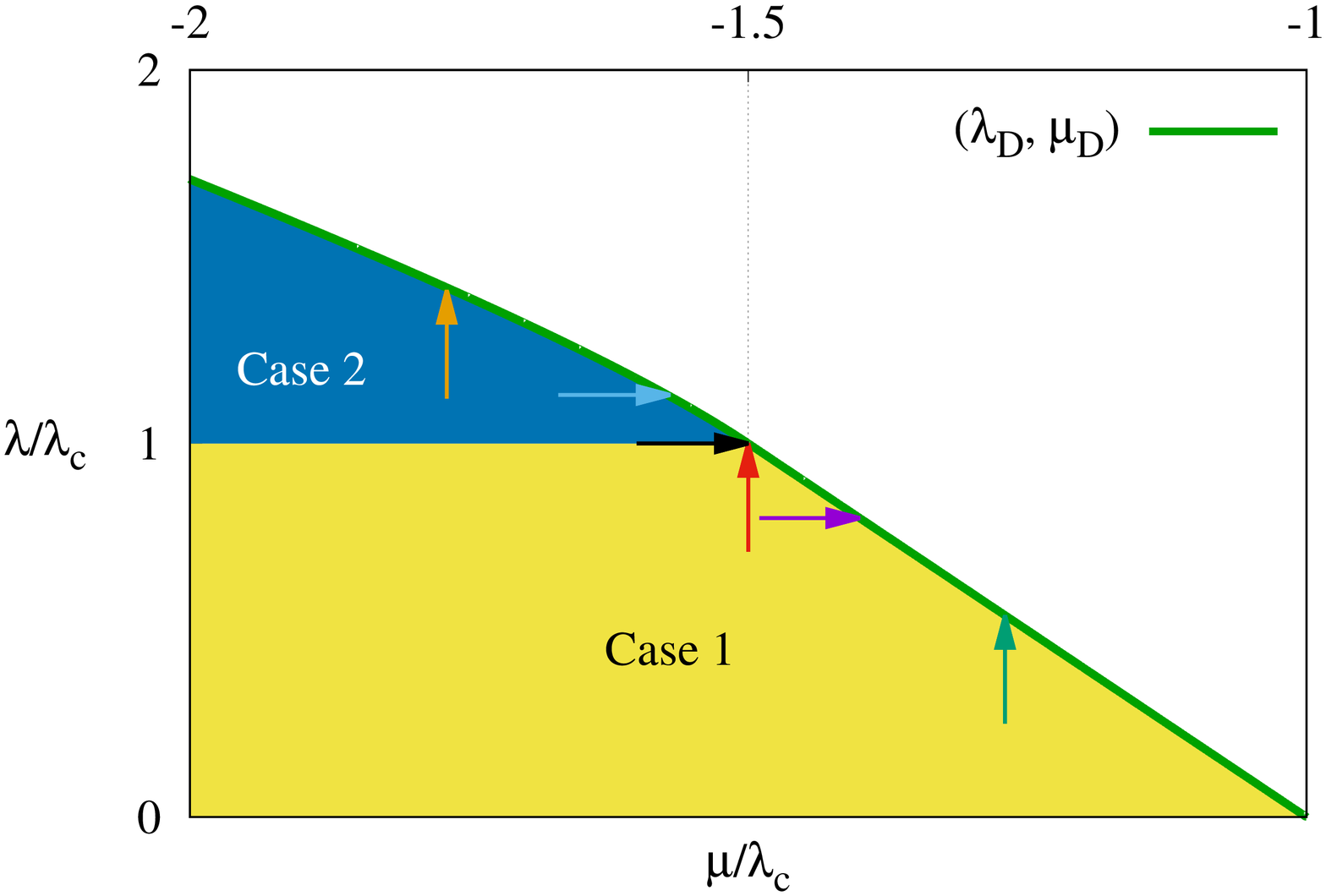}\includegraphics[width=0.35\textwidth,angle=-90]{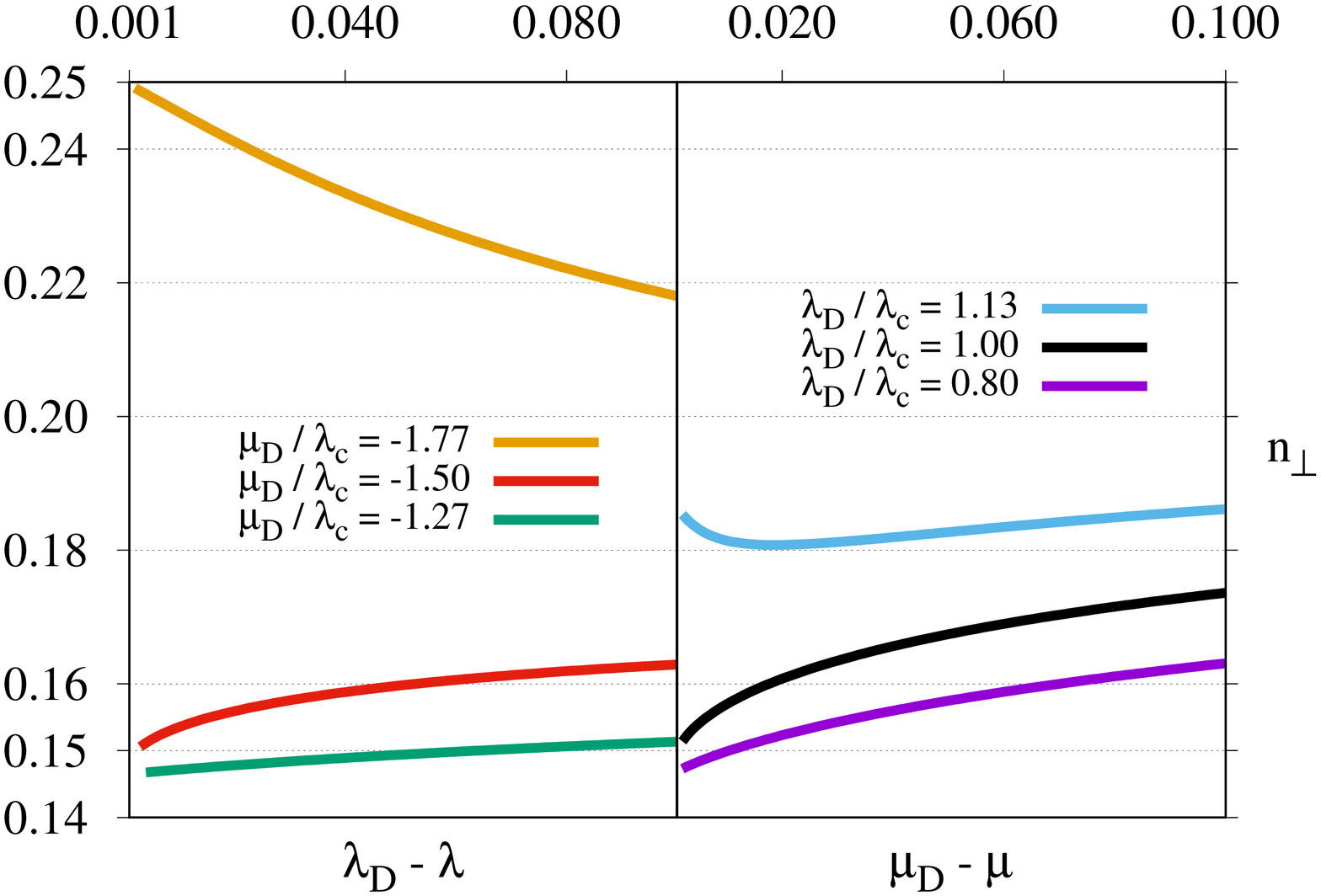}
\caption{
Left. Colored regions represent the parameter zones where the grand partition
function converged (equilibrium zones). Right. Relative mean number of particles in
the mode perpendicular to the mode with maximum occupation, as defined by equation
(\ref{e04151}). Perpendicular occupation falls when approaching the divergence from
inside the case-1 region or through the boundary separating the cases, showing that
to first order the particles tend to gather all in the mode with maximum
occupation.  This mode coincides with the ground state mode in the range $\lambda <
\lambda_c$.  The case is different when approaching the divergence from the
case-2 region, where the perpendicular occupation tends to a finite number. The
values $\gamma = 1$ and $\beta = 1$ were taken to produce the graphs at the right
panel.
}
\label{fig1}
\end{center}
\end{figure*}
\begin{table}
\begin{center}
\begin{tabular}{|c|c|} \hline 
 $\Xi$ & $ \xi \left[(\lambda_D - \lambda) \kappa(\lambda) + \mu_D - \mu \right]^{-\alpha} $
\\ \hline
$\lambda_c$ & $\sqrt{1+\gamma^2}$ \\ \hline
$\lambda_D$ & 
$\begin{array}{cc}
-2(\lambda_c + \mu_D) & \text{if } \mu_D  \ge -\frac{3}{2} \lambda_c, \  \\
\frac{1}{2}\left( -\mu_D + \sqrt{\mu_D^2 - 2 \lambda_c^2 } \right) & \text{if }
 \mu_D \le -\frac{3}{2} \lambda_c. 
\end{array}$ \\ \hline
$\kappa(\lambda)$ & 
$
\begin{array}{cc}
\frac{1}{2}                             & \text{if } \lambda_D \le \lambda_c, \\
 1 - \frac{\lambda_c^2}{2 \lambda \lambda_D}  & \text{if } \lambda_D \ge \lambda_c.
\end{array}$ \\ \hline
$\alpha$ & 
$\begin{array}{cc}
 1                             & \text{if } (\lambda_D,\mu_D) \ne
 (\lambda_c,-\frac{3}{2} \lambda_c), \\
 \frac{5}{4}                   & \text{if } (\lambda_D,\mu_D) =
 (\lambda_c,-\frac{3}{2} \lambda_c).
\end{array}$ \\ \hline
$\xi$ & 
\begin{tabular}{c}
A function of $\lambda$, $\mu$ and $\beta$ that tends to \\
a finite value when $(\lambda,\mu) \rightarrow (\lambda_D^-,\mu_D^-)$. 
\end{tabular} \\ \hline
\end{tabular}
\end{center}
\caption{Grand partition function of Hamiltonian (\ref{e04091}). Notice that
$\lambda<\lambda_D$ and $\mu<\mu_D$.}
\label{t04111}
\end{table}
\begin{table*}
\begin{center}
\begin{tabular}{|c|c|c|c|} \hline 
Observable & Formal expression & Leading term close to $(\lambda_D,\mu_D)$ &
Behavior near $(\lambda_D,\mu_D)$ $\left(\times \mathscr{M} \right)$ \\ \hline
$\mathscr{M} = \langle \hat{m}_1  + \hat{m}_2 \rangle$ & $\frac{1}{\beta}\frac{\partial \log \Xi
}{\partial \mu}  $ & $\frac{\alpha}{\beta} \left[ \kappa(\lambda_D)(\lambda_D -
\lambda) + \mu_D - \mu  \right ]^{-1} $ & $1$ \\ \hline
$E = \langle \hat{H} \rangle$ & $\mu \mathscr{M}
-\frac{\partial \log \Xi}{\partial \beta}$ & $\mu_D \mathscr{M} $
& 
$\begin{array}{cc}
-\lambda_c - \frac{\lambda_D}{2} & \text{if } \lambda_D  \le \lambda_c,   \\
-\lambda_D - \frac{\lambda_c^2}{2 \lambda_D}   & \text{if }
 \lambda_D \ge \lambda_c. 
\end{array}$ \\ \hline
$I = \langle  \frac{\hat{m}_1^2 + \hat{m}_2^2}{M} \rangle$ &  $\frac{1}{\beta} \frac{\partial \log
\Xi}{\partial \lambda}$ & $\kappa(\lambda_D) \mathscr{M} $
& 
$\begin{array}{cc}
\frac{1}{2} & \text{if } \lambda_D  \le \lambda_c, \  \\
1 - \frac{\lambda_c^2}{2 \lambda_D^2}   & \text{if } \lambda_D \ge \lambda_c. 
\end{array}$ \\ \hline
$J = - i \langle \hat{a}_1^{\dagger}\hat{a}_2 - \hat{a}_2^{\dagger} \hat{a}_1 \rangle$ &  $-\frac{1}{\beta} \frac{\partial \log
\Xi}{\partial \gamma}$ & $\frac{\gamma \kappa(\lambda_D)}{\lambda_c} \frac{\partial
\lambda_D}{\partial \lambda_c} \mathscr{M} $ & 
$\begin{array}{cc}
-\frac{\gamma}{\lambda_c} & \text{if } \lambda_D  \le \lambda_c, \  \\
-\frac{\gamma}{\lambda_D} & \text{if } \lambda_D \ge \lambda_c. 
\end{array}$ \\ \hline
$W = \langle \hat{a}_1^{\dagger}\hat{a}_2 + \hat{a}_2^{\dagger} \hat{a}_1 \rangle$ & $E -\gamma J 
 + \lambda I  $  & $ \left[ \mu_D +
\kappa(\lambda_D) \left( \lambda_D - \frac{\gamma^2 }{\lambda_c} \frac{\partial
\lambda_D}{\partial \lambda_c} \right) \right] \mathscr{M} $ & 
$\begin{array}{cc}
-\frac{1}{\lambda_c} & \text{if } \lambda_D  \le \lambda_c, \  \\
-\frac{1}{\lambda_D} & \text{if } \lambda_D \ge \lambda_c. 
\end{array}$ \\ \hline
\end{tabular}
\end{center}
\caption{Mean values according to the grand canonical formalism. Unreferenced
variables are identified in table \ref{t04111}.}
\label{t04112}
\end{table*}
When the system is embedded in a bath of inverse temperature $\beta$ and chemical
potential $\mu$ it eventually reaches thermodynamic equilibrium. This equilibrium
state is represented by a mixed state whose most important characterization is
given by the grand canonical partition function
\begin{gather}
\Xi = \sum_{M=0}^\infty Tr \left( e^{-\beta (\hat{H}_M - \mu M )} \right).
\end{gather}
The presence of non-quadratic terms in the Hamiltonian hinders the exact analytical
calculation of $\Xi$. Hence, the following quadratic form is proposed
\begin{widetext}
\begin{gather}
\Xi \approx 1 + \frac{1}{\sqrt{\pi}} \sum_{M=1}^\infty e^{\frac{\beta \lambda
M}{2}} \int_{-\infty}^{\infty} dx  e^{-x^2} Tr
\left(e^{-\beta(\hat{a}_1^{\dagger}\hat{a}_2 + \hat{a}_2^{\dagger} \hat{a}_1) + i
\beta \gamma (\hat{a}_1^{\dagger}\hat{a}_2 - \hat{a}_2^{\dagger} \hat{a}_1)  +
x\sqrt{\frac{2 \beta \lambda}{M}} (\hat{m}_1 - \hat{m}_2 )  + \beta \mu M }
\right).   
\label{e04092}
\end{gather}
\end{widetext}
The original grand partition function can be obtained from this expression by
carrying out the integral treating operators as scalars. This alternative
expression is expected to be accurate over parameter zones where the system
displays high occupation, since in this case fluctuations become negligible
compared to mean values. As can be seen from appendix \ref{a04192}, in the process
of calculating (\ref{e04092}) it is found that $\Xi$ converges inside the parameter
zones highlighted in figure \ref{fig1}. The grand partition function tends to
diverge as any point of the curve $(\lambda_D,\mu_D)$, whose explicit expression is
given in table \ref{t04111}, is approached from the left. Near this divergence,
$\Xi$ displays the functionality reported also in table \ref{t04111}. The line
$\lambda = \lambda_c$ partitions the parameter space into cases whose properties
are being discussed further ahead in this note. However, there is no symmetry
breaking. In both cases the number of particles at both wells is the same. This is
to be expected from the fact that the thermodynamic state is a function of the
Hamiltonian an as such it must display its symmetries. A number of observables can
be obtained as derivatives of the grand partition function. These can be seen in
table \ref{t04112}. They all show the same scaling behavior near divergence,
namely, proportional to the total number of particles.
However, multiplicative coefficients display different functionality depending on
the region of parameter space from which the divergence is approached. The relation
between any pair of observables is continuous, but the relation between an
observable and the divergence point presents a discontinuity at $(\lambda_D,\mu_D)
= (\lambda_c,-\frac{3}{2} \lambda_c)$. Thus, at finite temperature the parameter
space is divided into two cases that can be recognized according to their
divergence coefficients. 
\section{Population distribution}
Knowing that observables diverge when the system's parameters approach an
established boundary in parameter space, it is of interest to uncover some of the
features that characterize the thermodynamic state near this divergence zone. One
way of addressing this issue is to find how much a given mode contributes to the
thermodynamic state. Although bosons can access two independent modes, one for each
well in Hamiltonian (\ref{e04091}), it is possible to build other (linearly
dependent) modes as superpositions of the original ones. In this scenario it becomes
relevant to find the mean occupation of a given mode. For this purpose let us
define the following weight function
\begin{gather}
N(\theta,\varphi) =  \langle \hat{b}^\dagger \hat{b} \rangle \text{, }
\hat{b}^{\dagger} =  \hat{a}_1^{\dagger} \cos \theta - \hat{a}_2^{\dagger} e^{i
\varphi}\sin \theta.
\end{gather}
The average is calculated over the thermodynamic state associated with the
grand partition function previously introduced in this document. Any mode available
in physical space can be reached in the range $0\le\theta\le \pi/2$ and
$0\le\varphi<2\pi$. From a direct calculation it can be shown that this weight
function can also be written as
\begin{gather}
N = \frac{\mathscr{M}}{2} - \frac{\sin 2 \theta}{2} ( W \cos
\varphi  +  J \sin \varphi ).
\label{e01011}
\end{gather}
In the process of arriving at this expression the property $\langle \hat{m}_1 -
\hat{m}_2 \rangle = 0$ was used. This property follows from the fact that the
equilibrium state must display the Hamiltonian's inversion symmetry. An explicit
form of $N$ can be obtained by replacing the mean values of table \ref{t04112} in
(\ref{e01011}).  A pair of angles $(\theta^\blacktriangle,\varphi^\blacktriangle)$
defining a mode with maximum contribution must fulfill 
\begin{gather}
\left . \partial_\theta N \right |_{\theta^\blacktriangle \varphi^\blacktriangle} =
-\cos 2\theta^\blacktriangle  \left ( W \cos \varphi^\blacktriangle   +  J \sin
\varphi^\blacktriangle \right ) = 0 \label{e01012} \\ 
\left . \partial_\varphi N  \right |_{\theta^\blacktriangle \varphi^\blacktriangle}
=  \frac{\sin 2\theta^\blacktriangle}{2}  \left ( W \sin \varphi^\blacktriangle  -
J \cos \varphi^\blacktriangle \right ) = 0 \label{e01013} \\
\left . \partial_{\theta^2}^2 N \right |_{\theta^\blacktriangle
\varphi^\blacktriangle}  = 2 \sin 2\theta^\blacktriangle  \left ( W \cos
\varphi^\blacktriangle  +  J \sin \varphi^\blacktriangle \right ) < 0.
\label{e01014} \\
4 \left . \partial_{\varphi^2}^2 N  \right |_{\theta^\blacktriangle
\varphi^\blacktriangle} = \left . \partial_{\theta^2}^2 N \right
|_{\theta^\blacktriangle \varphi^\blacktriangle} < 0 .\label{e01015} 
\end{gather}
As can be seen, condition (\ref{e01014}) implies (\ref{e01015}), so it is really
three conditions that must be accounted for. Equation (\ref{e01012}) is met for
$\theta^\blacktriangle = \pi/4$, giving a state with equal occupation at
both sides, in compliance with the Hamiltonian's symmetry. According to table
\ref{t04112} both $W$ and $J$ are negative in the zone near divergence, hence
conditions (\ref{e01013}) and (\ref{e01014}) are satisfied for 
{\small
\begin{gather}
\cos \varphi^\blacktriangle = \frac{1}{\sqrt{1+t^2}}, \text{ } \sin
\varphi^\blacktriangle = \frac{t}{\sqrt{1+t^2}}, \text{ } t = \frac{J}{W} = \gamma.
\end{gather}
}
Interestingly, these parameters also define the ground state in the gaped phase
($\lambda<\lambda_c$) of the quantum phase transition discussed in the first
part of this work, but in this case this state with maximum occupation applies
to the whole spectrum of parameters near divergence, including the gapless
phase. The maximum occupation number is given by
\begin{gather}
N^\blacktriangle = N(\theta^\blacktriangle,\varphi^\blacktriangle) = \frac{\mathscr{M}
}{2} + \frac{1}{2} \sqrt{W^2 + J^2}. 
\end{gather}
Inserting the values reported in table \ref{t04112} it results
\begin{gather}
n = \frac{N^\blacktriangle}{\mathscr{M}} = \begin{cases} 1  & \text{if } \lambda_D \le \lambda_c,
\\ \frac{1}{2} \left( 1 + \frac{\lambda_c}{\lambda_D}  \right ) &
\text{if } \lambda_D \ge \lambda_c.  \end{cases}
\end{gather}
As below $\lambda_c$ the number of particles in the most populated mode
coincides with the mean number of particles in the whole system, it is argued
that in this regime particles go all into the maximally occupied mode, which is
at the same time the ground state of the high-density Hamiltonian. The quantity
\begin{gather}
n_{\bot} = 1  - n,
\label{e04151}
\end{gather}
is also the relative mean number of particles in the mode perpendicular to
$\hat{b}^\dagger$, $\hat{d}^\dagger = \hat{a}_1^{\dagger} \cos \theta +
\hat{a}_2^{\dagger} e^{i \varphi}\sin \theta$, evaluated at
$(\theta^\blacktriangle,\varphi^\blacktriangle)$. It then follows that for the
case-1 family of parameters of figure \ref{fig1} the number of particles in the
maximally occupied state approaches the mean number of particles in the system at
the same time that the number of particles in the respective perpendicular state
becomes negligible. Instead, in the case-2 region the population is distributed
over two modes.  Take into account that this derivation is valid near the curve of
divergence and not necessarily over the whole space of parameters.

A numerical study is undertaken in order to benchmark these analytical results.
Hamiltonian (\ref{e04091}) is diagonalized for various system sizes $M$ and the
respective eigenvalues $E_j^{M}$ are used to find the grand canonical partition
function thus
\begin{gather}
\Xi = 1 + \sum_{M=1}^{M_{ax}} e^{\beta \mu M} \sum_{j=1}^{M+1} e^{-\beta E_j^M}.
\label{e04152}
\end{gather}
The maximum size $M_{ax}$ is adjusted depending on the system's parameters to
achieve an absolute accuracy of $10^{-7}$. Eigenvalues can be in addition employed
to calculate the mean number of particles $\mathscr{M}$ using a similar expression.
Eigenvectors are also calculated and used to find $J$ and $W$ as weighed sums (not
as numerical derivatives) in similarity to (\ref{e04152}). These values are then
used to find $n_{\bot}$. The results are shown in the right panel of figure
\ref{fig1}. The numerical values of $n_{\bot}$ show a tendency that is consistent
with the analytical study, displaying a monotonously decaying pattern as the
divergence is approached by either $\lambda$ or $\mu$ from the case-1 region. The
same tendency is observed when the divergence is approached through the line that
divides the parameter space. The decaying is stronger when approaching
$(\lambda_D,\mu_D) = (\lambda_c,-\frac{3}{2} \lambda_c)$ (black and red curves),
scaling as $\approx x^{0.02}$ close to the origin. In other cases the decaying
coefficient is smaller and seems to be dependent on the divergence point.
Furthermore, $n_{\bot}$ grows when approaching $\lambda_D$ or $\mu_D$ at close
range from the case-2 region, suggesting a macroscopic fraction of particles settle
in the perpendicular mode. These features show that the properties of the
high-density ground state dominate the collective response in thermodynamic
equilibrium \cite{zhang}.  The gaped phase transmutes into a collective state in
which bosons cluster together in a single mode. This behavior seems to be
independent of temperature, rather being determined by the proximity of interaction
and chemical potential to the divergence region. Nevertheless, it is likely that
temperature would define how fast collectivism arises as the divergence is
approached.  The fact that compresibility, which can be found as $\partial_\mu
\mathscr{M}$, goes to infinity in general as $(\lambda,\mu) \rightarrow
(\lambda_D,\mu_D)$ is indicative of a highly conductive state not necessarily
correlated with collectivism. 

A condensate is traditionally understood as the macroscopic occupation of a
single-body state that in normal conditions displays only marginal occupation, like
any other state in the spectrum. Here the situation is seen from the perspective
that two occupation states are initially macroscopically occupied and then one of
them is depleted under the right conditions. The end result is the same if the
depleted state is seen as the set of marginally occupied states in the condensate.
\section{Conclusions}
The equilibrium response of a bosonic system displaying a quantum phase transition
has been studied using the grand canonical formalism. The phases characterizing the
zero-temperature state drive characteristic features in the finite temperature
regime near the parameter zone where the partition function diverged. A form of
quantum collectivism in which occupation is concentrated in the ground-state
mode has been identified. Too much attraction tends to dissolve this state, but
there exist a critical value of the interaction constant below which collectivism
can prevail in the presence of attractive interaction. This is related to the
effect that the negative chemical potential considered here has on balancing the
instability caused by the attractive interaction and the fact that the interaction
constant has been rescaled with respect to the system size.  Moreover, according to
the phase diagram in figure \ref{fig1} there cannot be a condensed phase for
positive values of the chemical potential.  Perhaps the most critical aspect of the
present study is therefore the consideration of negative chemical potential, which
might pose technical challenges in a controlled scenario \cite{tobias}, but could
take place spontaneously in a less-artificial one, as for example a bosonic
superfluid.  It remains to be seen whether other forms of collectivism can be
induced by deliberately breaking the system's inversion symmetry, for instance, by
setting different chemical-potential variables at each well. Also of interest is to
scale up the model by considering a large number of modes in one dimension and see
whether the collectivism observed in the small system can be a precursor of
condensation in an infinite chain. Ultimately, by adding bosonic modes describing
molecules it should be possible to study the interplay between attractiveness and
molecular formation.
\appendix
\section{Quantum Phase Transition}
\label{a04191}
Making $\delta=1$ Hamiltonian (\ref{e04091}) becomes
\begin{gather}
\hat{H}_M = \hat{a}_1^{\dagger}\hat{a}_2 + \hat{a}_2^{\dagger} \hat{a}_1 - i \gamma
(\hat{a}_1^{\dagger}\hat{a}_2 - \hat{a}_2^{\dagger} \hat{a}_1) \nonumber \\
- \frac{\lambda}{M} \left ( \hat{a}_1^{\dagger}\hat{a}_1
\hat{a}_1^{\dagger}\hat{a}_1 + \hat{a}_2^{\dagger}\hat{a}_2
\hat{a}_2^{\dagger}\hat{a}_2 \right ).  \label{april81}
\end{gather}
It is assumed that the ground state can be written as
\begin{gather}
|G (\theta,\varphi) \rangle = \frac{\left. {\hat{b}_1^{\dagger}}\right.^M
|0,0\rangle}{\sqrt{M!}} = |M,0\rangle,  \\ 
\hat{b}_1^{\dagger} =  \hat{a}_1^{\dagger} \cos \theta-  \hat{a}_2^{\dagger} e^{i
\varphi}\sin \theta,
\label{april82}
\end{gather}
so that $[\hat{b}_1,\hat{b}_1^\dagger]=1$. The angle domains are
$0\le\theta\le\frac{\pi}{2}$ and $0\le\varphi<2\pi$, in such a way that different
angle pairs correspond to different modes.  The basis change is given by
\begin{gather}
\left [
\begin{array}{c}
\hat{b}_1^\dagger \\
\hat{b}_2^\dagger 
\end{array}
\right ] =
\left [
\begin{array}{cc}
\cos \theta & -e^{i \varphi} \sin \theta \\
e^{-i \varphi} \sin \theta & \cos \theta
\end{array}
\right ] 
\left [
\begin{array}{c}
\hat{a}_1^\dagger \\
\hat{a}_2^\dagger 
\end{array}
\right ] \\
\Rightarrow \left [
\begin{array}{c}
\hat{a}_1^\dagger \\
\hat{a}_2^\dagger 
\end{array}
\right ] =
\left [
\begin{array}{cc}
\cos \theta & e^{i \varphi} \sin \theta \\
-e^{-i \varphi} \sin \theta & \cos \theta
\end{array}
\right ] 
\left [
\begin{array}{c}
\hat{b}_1^\dagger \\
\hat{b}_2^\dagger 
\end{array}
\right ].
\label{april83}
\end{gather}
From direct calculations the following results are obtained
\begin{gather}
\langle G  | \hat{a}_1^{\dagger}\hat{a}_2   | G 
\rangle = - M e^{i \varphi} \cos \theta \sin \theta, \\
\langle G  | \hat{a}_1^{\dagger}\hat{a}_1
\hat{a}_1^{\dagger}\hat{a}_1  | G  \rangle = M^2 \cos^4\theta + M
\cos^2\theta \sin^2\theta,  \\
\langle G  | \hat{a}_2^{\dagger}\hat{a}_2
\hat{a}_2^{\dagger}\hat{a}_2  | G  \rangle = M^2 \sin^4\theta + M
\cos^2\theta \sin^2\theta. 
\end{gather}
Using these values the energy becomes to leading order in $M$
\begin{gather}
E_G(\theta,\varphi) = \langle G (\theta,\varphi) | \hat{H}_M G (\theta,\varphi)
\rangle = \nonumber \\ -M \left ( \sin 2 \theta (\cos \varphi + \gamma \sin \varphi) + \lambda \left( 1 - \frac{\sin^2 2 \theta}{2} \right) \right ).
\label{april85}
\end{gather}
The ground state energy corresponds to the minimum of this expression with
respect to $\theta$ and $\varphi$. Defining
\begin{gather}
q(\theta,\varphi) = \sin 2 \theta (\cos \varphi + \gamma \sin \varphi) - \frac{\lambda \sin^2 2 \theta}{2},
\label{april86}
\end{gather}
extreme points must satisfy
{\small
\begin{gather}
\left . \partial_{\theta} q  \right |_{\theta^\star  \varphi^\star}  =  \cos 2
\theta^\star ( \cos \varphi^\star + \gamma \sin \varphi^\star - \lambda \sin 2
\theta^\star ) = 0, \label{april91} \\ 
\left . \partial_{\varphi} q \right |_{\theta^\star  \varphi^\star}  = -\sin 2 \theta^\star (\sin \varphi^\star - \gamma \cos \varphi^\star) = 0, \label{april92} \\
\left . \partial_{\theta^2}^2 q \right |_{\theta^\star  \varphi^\star}  = -2 \sin 2
\theta^\star ( \cos \varphi^\star + \gamma \sin \varphi^\star - \lambda \sin 2
\theta^\star ) \nonumber \\
- \lambda \cos^2 2 \theta^\star < 0, \label{april93} \\
\left . \partial_{\varphi^2}^2 q \right |_{\theta^\star  \varphi^\star}  = \sin 2 \theta^\star (-\cos \varphi^\star - \gamma \sin \varphi^\star) < 0. \label{april94} 
\end{gather}
}
Conditions (\ref{april91}), (\ref{april92}) and (\ref{april94}) are all met for the
next values
\begin{gather}
\theta^\star = \frac{\pi}{4}, \text{ } \cos \varphi^\star = \frac{1}{\sqrt{1+\gamma^2}}, \text{ } \sin \varphi^\star = \frac{\gamma}{\sqrt{1+\gamma^2}}.
\end{gather}
Condition (\ref{april93}) is met for these same values in the range $\lambda <
\sqrt{1 + \gamma^2}$, which defines the scope of this particular physical phase.
Replacing these extreme points in (\ref{april85}) the ground state energy in this
phase is found to be
\begin{gather}
E_G = -M \left( \sqrt{1+\gamma^2} + \frac{\lambda}{2} \right ).
\end{gather}
Correspondingly, over the range $\lambda > \sqrt{1 + \gamma^2}$ the following
values constitute solutions
\begin{gather}
\theta_1^\star = \frac{1}{2} \arcsin \frac{\sqrt{1+\gamma^2}}{\lambda}, \text{ }
\theta_2^\star =  \frac{\pi}{2} - \theta_1^\star, \\ 
\cos \varphi^\star = \frac{1}{\sqrt{1+\gamma^2}}, \text{ } \sin \varphi^\star = \frac{\gamma}{\sqrt{1+\gamma^2}}.
\end{gather}
The solutions are given by the pairs $\{\theta_1^\star,\varphi^\star\}$ and
$\{\theta_2^\star,\varphi^\star\}$. Either pair delivers the following ground state
energy
\begin{gather}
E_G = -M \left( \lambda + \frac{1+\gamma^2}{2 \lambda} \right ).
\end{gather}
The expression
\begin{gather}
\lambda_c = \sqrt{1+\gamma^2} 
\end{gather}
is the transition's critical point. Since these results correspond to first order
in $M$, the quantum phase transition takes place in the high density limit, i.e.,
$M \rightarrow \infty$.
\section{Calculation of $\Xi$}
\label{a04192}
Taking equation (\ref{e04092}) and making the variable change $x = y \sqrt{M
\beta}$ leads to the next expression
{\small
\begin{gather}
\Xi \approx 1 + \sqrt{\frac{\beta}{\pi}} \sum_{M=1}^\infty
\sqrt{M} e^{\frac{\beta \lambda M}{2}} \int_{-\infty}^{\infty} dy  e^{- \beta M  y^2} Tr(e^{\beta \hat{h}_M}),   
\label{abril232}
\end{gather}
}
so that
\begin{gather}
\hat{h}_M = -(\hat{a}_1^{\dagger}\hat{a}_2 + \hat{a}_2^{\dagger} \hat{a}_1) + i
\gamma (\hat{a}_1^{\dagger}\hat{a}_2 - \hat{a}_2^{\dagger} \hat{a}_1) \nonumber \\
+ \sqrt{2 \lambda} y (\hat{m}_1 - \hat{m}_2 )  + \mu (\hat{m}_1 + \hat{m}_2).
\end{gather}
This effective Hamiltonian can also be written as
\begin{gather}
\hat{h}_M =
(\hat{a}_1^\dagger \text{ } \hat{a}_2^\dagger)
\left (
\begin{array}{cc}
 \mu  + \sqrt{2 \lambda} y  & -1 + i \gamma \\
-1 - i \gamma  &  \mu -\sqrt{2 \lambda} y
\end{array}
\right )
\left(
\begin{array}{c}
\hat{a}_1 \\
\hat{a}_2
\end{array}
\right).
\end{gather}
Normal eigenenergies of $\hat{h}_M$ are then found to be
\begin{gather}
\epsilon_{\pm}(y) = \mu \pm \sqrt{ 2 \lambda y^2 + 1 + \gamma^2 } = \mu \pm \sqrt{
2 \lambda y^2 + \lambda_c^2 }.
\end{gather}
Using these values the trace can be calculated as follows
\begin{gather}
Tr(e^{\beta \hat{h}_M}) = Tr(e^{\beta ( \epsilon_+ \hat{b}_+^\dagger \hat{b}_+ +
\epsilon_- \hat{b}_-^\dagger \hat{b}_-)}) \nonumber \\
= \sum_{n=0}^M e^{\beta \epsilon_+ n +
\beta \epsilon_-(M-n)} = \frac{e^{\beta M\epsilon_-} - e^{\beta(\epsilon_+ - \epsilon_-)}
e^{\beta M\epsilon_+}}{1 - e^{\beta(\epsilon_+ - \epsilon_-)}}.
\label{abril234}
\end{gather}
Operators $\hat{b}_+$ and $\hat{b}_-$ correspond to diagonal bosonic modes.  Now
let us take a term from (\ref{abril232}) and reformulated in the next
way
\begin{gather}
\sqrt{M} e^{\frac{\beta \lambda M}{2}} = \frac{\sqrt{\beta} M}{\sqrt{\pi}}
\int_{-\infty}^{\infty} dx e^{- \beta M x^2 + \beta M \sqrt{2 \lambda} x}.
\label{abril235}
\end{gather}
Replacing (\ref{abril234}) and (\ref{abril235}) in (\ref{abril232}) and organizing
terms yields
{\scriptsize
\begin{gather}
\Xi = 1 + \frac{\beta}{\pi} \int_{-\infty}^{\infty} dx \int_{-\infty}^{\infty} dy
\frac{ \sum_{M=1}^\infty M e^{\beta M ( -x^2-y^2 + \sqrt{2 \lambda}x +
\epsilon_+(y))}}{1-e^{-\beta(\epsilon_+(y) - \epsilon_-(y))}}  \nonumber \\
+ \frac{\sum_{M=1}^\infty M e^{\beta M ( -x^2-y^2 + \sqrt{2 \lambda}x + \epsilon_-(y))}}{1-e^{\beta(\epsilon_+(y) - \epsilon_-(y))}} .  
\label{abril237}
\end{gather}
}
Defining
\begin{gather}
F(x,y) = -x^2-y^2 + \sqrt{2 \lambda}x + \sqrt{ 2 \lambda y^2 + \lambda_c^2 } + \mu,
\end{gather}
it can be seen that both sums in (\ref{abril237}) shall converge if $F(x,y)<0$ for
any real value of $x$ and $y$. For this it is necessary that
\begin{gather}
F(x^{\star},y^{\star}) = F^\star <0,
\end{gather}
where $x^{\star}$ and $y^{\star}$ demark the location of the function's global
maximum. Through a calculus analysis it can be shown that two main cases arise  
\subsection{\bf {case 1:}  $0 \le \lambda < \text{min}(-2 ( \mu + \lambda_c),
\lambda_c)$.}
The function's only maximum is located at $ (x^\star = \sqrt{\lambda/2}, y^\star =
0 )$. The system's parameters are compatible with the condition
\begin{gather}
F^{\star} = \frac{\lambda}{2} + \lambda_c + \mu = -\frac{(\lambda_D - \lambda)}{2}  - (\mu_D-\mu)  < 0,
\label{e12101}
\end{gather}
being $\lambda_D$ and $\mu_D$ a pair of constants satisfying  $\lambda_D =
-2(\lambda_c + \mu_D)$.  Equation (\ref{e12101}) highlights the fact that
\begin{gather}
\lim_{\lambda \rightarrow \lambda_D^-,\mu \rightarrow \mu_D^-} F^{\star} = 0.
\label{pandora}
\end{gather}
\subsection{\bf{case 2:} $\frac{\mu}{\lambda_c} < -\frac{3}{2}$ and $\lambda_c \le
\lambda < \frac{1}{2}\left( -\mu + \sqrt{\mu^2 - 2 \lambda_c^2 } \right)$.}
The function displays two maxima located at 
\begin{gather}
\left( x^\star =
\sqrt{\frac{\lambda}{2}}, y^\star = \sqrt{\frac{\lambda^2 - \lambda_c^2}{2
\lambda}} \right),   
\end{gather}
and
\begin{gather}
\left( x^\star =
\sqrt{\frac{\lambda}{2}}, y^\star = -\sqrt{\frac{\lambda^2 - \lambda_c^2}{2 \lambda}} \right).
\end{gather}

The system's parameters are compatible with the condition
\begin{gather}
F^{\star} = \lambda + \frac{\lambda_c^2}{2 \lambda} + \mu \nonumber \\
= -(\lambda_D - \lambda) \left( 1 - \frac{\lambda_c^2}{2 \lambda \lambda_D}
\right) - (\mu_D - \mu) < 0,
\label{e11082}
\end{gather}
such that $\lambda_D = \frac{1}{2}\left( -\mu_D + \sqrt{\mu_D^2 - 2 \lambda_c^2 }
\right)$. As in the previous case, equation (\ref{pandora}) is satisfied. 
Figure \ref{fig1} depicts the two cases in a parameter map. The divergence
parameters, $(\lambda_D,\mu_D)$, are related as follows
{\small
\begin{gather}
\lambda_D = 
\begin{cases}
-2(\lambda_c + \mu_D) & \text{if } \mu_D  \ge -\frac{3}{2} \lambda_c, \  \\
\frac{1}{2}\left( -\mu_D + \sqrt{\mu_D^2 - 2 \lambda_c^2 } \right) & \text{if }
 \mu_D \le -\frac{3}{2} \lambda_c. 
\end{cases}
\end{gather}
}
Solving the sums in (\ref{abril237}) within the established spaces of convergence
the following result is obtained
\begin{gather}
\Xi \approx 1 + \frac{\beta}{4 \pi } \int_{-\infty}^{\infty} dx
\int_{-\infty}^{\infty} dy \frac{\text{csch}^2 \left( \frac{\beta}{2}F(x,y)
\right)}{1-e^{-\beta \surd (y)}} \nonumber \\
+ \frac{\text{csch}^2\left( \frac{\beta}{2} G(x,y) \right )}{1-e^{\beta \surd(y)}},
\label{e10301}
\end{gather}
where
{\small
\begin{gather}
G(x,y) =  F(x,y) - \surd(y), \text{ } \surd(y) = 2 \sqrt{ 2 \lambda y^2 + \lambda_c^2} .
\end{gather}
}
Close to divergence, only the part of the integral with $F(x,y)$ in (\ref{e10301})
goes to infinity. Since in such a case $F(x,y)$ gets close to zero, the following
approximation becomes applicable
\begin{gather}
\int_{-\infty}^{\infty} dx \int_{-\infty}^{\infty} dy \frac{\text{csch}^2 \left(
\frac{\beta}{2}F(x,y) \right)}{1-e^{-\beta \surd(y)}}  \nonumber \\
\approx  \frac{4}{\beta^2} \int_{-\infty}^{\infty} dx \int_{-\infty}^{\infty} dy \frac{e^{\beta F(x,y)}}{F(x,y)^2(1-e^{-\beta \surd(y)})}
\end{gather}
Neither exponential in the integrand has a significant contribution to the scaling
pattern of the gran-partition function. Therefore, the integral is further approximated by
\begin{gather}
\frac{4 e^{\beta F^\star}}{\beta^2  (1-e^{-\beta \surd(y^\star)})} \int_{-\infty}^{\infty} dx \int_{-\infty}^{\infty} dy \frac{1}{ F(x,y)^2}.
\end{gather}
The integral above can be approximated close to the divergence zone, but for this
it is necessary to differentiate a number of subcases.
\subsection{\bf {case 1:}  $0 \le \lambda < \text{min}(-2 ( \mu + \lambda_c))$}
\subsubsection{\bf {subcase 1}: excluding $(\lambda_D, \mu_D) = ( \lambda_c,
-\frac{3 \lambda_c}{2})$}
Function $F(x,y)$ is expanded and the first nonvanishing terms are retained.
The resulting integral can be solved by standard methods, along the lines of
\begin{gather}
\int_{-\infty}^{\infty} dx \int_{-\infty}^{\infty} dy \frac{1}{ F(x,y)^2} \approx
\nonumber \\
\int_{-\infty}^{\infty} dx \int_{-\infty}^{\infty} dy \frac{1}{ \left( F^\star -
\alpha_x^2 (x-x^\star)^2 - \alpha_y^2 (y-y^\star)^2  \right)^2} \nonumber \\
= -\frac{\pi}{|\alpha_x| |\alpha_y| F^\star},
\label{e01111}
\end{gather}
being
{\small
\begin{gather}
\alpha_x^2 = - \frac{1}{2} \left . \frac{\partial^2 F}{\partial x^2} \right
|_{x=x^\star, y = y^\star},
\alpha_y^2 = - \frac{1}{2} \left . \frac{\partial^2 F}{\partial y^2} \right |_{x=x^\star, y =
y^\star}.
\end{gather}
}
An approximated expression to the grand partition function can be obtained by
gathering all the coefficients involved in the derivation. However, it is only
the denominator in the last term of (\ref{e01111}) that determines the
divergence trend. The resulting expression is valid inside the parameter zone
corresponding to this case. Replacing the following identities in (\ref{e01111}):
\begin{gather}
F^\star = -\frac{(\lambda_D - \lambda)}{2}  - (\mu_D-\mu), \\ 
\alpha_x^2 = 1,  \\ 
\alpha_y^2 = 1 - \frac{\lambda}{\lambda_c}, \label{e02111}
\end{gather}
the grand-partition function can be written as
\begin{gather}
\Xi = \frac{\xi(\lambda,\mu,\lambda_c)}{ \frac{\lambda_D - \lambda}{2} + \mu_D-\mu }.
\end{gather}
Function $\xi(\lambda,\mu,\lambda_c)$ must be well behaved and continuous at the
point $(\lambda,\mu) = (\lambda_D,\mu_D)$. The reason why this derivation cannot
accommodate $\lambda_D = \lambda_c$ is because this would allow $\lambda$ to get
infinitesimally close to $\lambda_c$, causing the vanishing of $\alpha_y$ in
(\ref{e02111}).
\subsubsection{ {\bf subcase 2}: $( \lambda_D, \mu_D ) =  \left( \lambda_c, -\frac{3 \lambda_c}{2} \right)$.} 
As one of the second-order expansion-terms of $F(x,y)$ vanishes when
$\lambda$ approaches $\lambda_c$, the next non-vanishing term of the expansion is
considered in (\ref{e01111}). The resulting expression reads
\begin{gather}
\int_{-\infty}^{\infty} dx \int_{-\infty}^{\infty} dy \frac{1}{ \left( F^\star -
\alpha_x^2 (x-x^\star)^2 - \gamma_y^2 (y-y^\star)^4  \right)^2} = \nonumber \\
\frac{1}{|\alpha_x| \sqrt{|\gamma_y|} (-F^\star)^{\frac{5}{4}}} \int_0^\pi
\frac{d\theta}{\sqrt{\sin \theta}} \int_0^\infty dt \frac{\sqrt{t}}{(1+t^2)^2}.
\label{e11083}
\end{gather}
Replacing
\begin{gather}
F^\star = -\frac{(\lambda_c - \lambda)}{2}  - (-\frac{3 \lambda_c}{2} -\mu), \\ 
\alpha_x^2 = 1, \\
\gamma_y^2 = \frac{\lambda^2}{2 \lambda_c^3},
\end{gather}
As a consequence, the grand partition function adopts the next form
\begin{gather}
\Xi = \frac{\xi(\lambda,\mu,\lambda_c)}{ \left( \frac{\lambda_c - \lambda}{2} -\frac{3 \lambda_c}{2} -\mu \right )^{\frac{5}{4}} },
\label{e11084}
\end{gather}
in such a way that $\xi(\lambda,\mu,\lambda_c)$ be well behaved and continuous at
$(\lambda,\mu) = (\lambda_c,-\frac{3 \lambda_c}{2})$.
\subsection{\bf{case 2:} $\frac{\mu}{\lambda_c} < -\frac{3}{2}$ and $\lambda_c \le
\lambda < \frac{1}{2}\left( -\mu + \sqrt{\mu^2 - 2 \lambda_c^2 } \right)$.} 
\subsubsection{ \bf {subcase 3}: excluding $ ( \lambda_D, \mu_D ) = \left(
\lambda_c,-\frac{3 \lambda_c}{2} \right)$.} 
In this case function $F(x,y)$ has two maxima located at opposite sides of the $y$
axis. Formally, this would require to consider the contribution of two expansions,
deriving in two integrals, each of which limited to half the plane. However, due to
the symmetry of $F(x,y)$, this is effectively equivalent to considering twice one
expansion integrated over the whole plane, since only the contribution around the
expansion point is relevant. Proceeding in this way, the same expression
(\ref{e01111}) found before is obtained. Likewise, replacing
\begin{gather}
F^\star =  -(\lambda_D - \lambda)\left( 1 - \frac{\lambda_c^2}{2 \lambda_D^2}  \right) - (\mu_D - \mu), \label{e11081} \\ 
\alpha_x^2 = 1, \\
\alpha_y^2 =  1 -\left( \frac{\lambda_c}{\lambda} \right )^2,
\end{gather}
it follows that the grand partition function can be written as
\begin{gather}
\Xi = \frac{\xi(\lambda,\mu,\lambda_c)}{(\lambda_D - \lambda)\left( 1 -
\frac{\lambda_c^2}{2 \lambda \lambda_D}  \right) + (\mu_D - \mu)},
\end{gather}
being $\xi(\lambda,\mu,\lambda_c)$ a well behaved and continuous function at
$(\lambda,\mu) = (\lambda_D,\mu_D)$. 
\subsubsection{ {\bf subcase 4}: $ ( \lambda_D, \mu_D ) = \left( \lambda_c,
-\frac{3 \lambda_c}{2} \right) $.} 
In close parallel to subcase 2, function $F(x,y)$ is expanded and the first
non-vanishing terms that survive in the limit $\lambda \rightarrow \lambda_c$ are
retained. The resulting expression coincides with (\ref{e11083}). Replacing
\begin{gather}
F^\star = -\frac{\lambda_c - \lambda}{2} - \left( -\frac{3 \lambda_c}{2} - \mu
\right), \\ 
\alpha_x^2 = 1, \\
\gamma_y^2 =  \frac{\lambda_c^2}{2 \lambda} \left( \lambda_c^2 - 4 (\lambda^2 -
\lambda_c^2)  \right),
\end{gather}
the final form of the grand-partition function is found to be analogous to
(\ref{e11084}). 
The grand partition function can be written in a general way as
\begin{gather}
\Xi = \frac{\xi}{\left[(\lambda_D - \lambda) \kappa(\lambda) + \mu_D - \mu
\right]^{\alpha}}.
\label{e11085}
\end{gather}
Function $\xi$ does not contribute to the scaling pattern of $\Xi$ and makes no
significant contribution to observables in the parameter zone near divergence. The
other variables are given by
\begin{gather}
\kappa(\lambda) = 
\begin{array}{cc}
\begin{cases}
\frac{1}{2}                             & \text{if } \lambda_D \le \lambda_c, \\
 1 - \frac{\lambda_c^2}{2 \lambda \lambda_D}  & \text{if } \lambda_D \ge \lambda_c.
\end{cases}
\end{array} 
\end{gather}
and
\begin{gather}
\alpha = 
\begin{array}{cc}
\begin{cases}
 1                             & \text{if } (\lambda_D,\mu_D) \ne
 (\lambda_c,-\frac{3}{2} \lambda_c), \\
 \frac{5}{4}                   & \text{if } (\lambda_D,\mu_D) =
 (\lambda_c,-\frac{3}{2} \lambda_c).
\end{cases}
\end{array}
\end{gather}
\end{document}